\documentstyle[aps,prl,epsfig,multicol]{revtex}
\begin{document}
\title{Singular Band Behavior of the Extended Emery Model for High-T$_c$ Superconductors}
\author{Ivana Mrkonji{\' c} and Slaven Bari{\v s}i{\' c}}
\address{Department of Physics, University of Zagreb, Bijeni{\v c}ka 32, POB 331, 10002 Zagreb,
Croatia}
\date{February 20, 2001.}
\maketitle
\begin{abstract}
The three band structure of the extended Emery model for the copper-oxide layered materials is
analyzed
for all the values of the effective tight-binding parameters. The model is characterized by the
Cu-O
site energy splitting, the Cu-O first neighbor hopping and the O-O second neighbor hopping. For the
sufficiently large O-O hoppings, the two bands can come together at one point on the diagonal of
the
square Brillouin zone, but they cannot cross. The range of the parameters for which the band
touching occurs is determined. The model relates the band touching at the bottom of the two bands
to the appearance of the extended, one-dimensional-like van Hove singularity at the lower edge of
the density of states. With $1+\delta$ holes ($\delta$ small) the extended van Hove singularity
can
thus occur close to the Fermi level only if the latter cuts two bands. The topology of the Fermi
surfaces observed by ARPES in the high-T$_c$ superconductors elliminates such possibility. The
Fermi surfaces in LSCO, Bi2212 and Y123 are then fitted with the one of the three bands at the
Fermi level. The agreement is excellent considering in particular the fact that the fits use only
two parameters. The departure of the measured band structure from the three band prediction found
on the small energy scales away from the Fermi level is attributed to the effects of the
fluctuations not included into the energy scales which define the three band structure.
\end{abstract}
\begin{multicols}{2}
\section{Introduction}
ARPES measurements of YBa$_2$Cu$_3$O$_{6.95}$ (Y123) \cite{Shabel1,Shabel2}, YBa$_2$Cu$_4$O$_8$
(Y124) \cite{Gofron}, 
Bi$_2$Sr$_2$CaCu$_2$O$_{8+\delta}$ (Bi2212)
\cite{Aebi,Ding,Mesot1,Dessau,Fretwell,Borisenko} and
La$_{2-x}$Sr$_x$CuO$_4$ (LSCO) \cite{Ino1,Ino2} Fermi surfaces provide interesting information
about the qualitative properties of the electron system in the high T$_c$ materials. At least
roughly, the electrons behave as fermions with large Fermi surfaces which pass through the vicinity
of the hyperbolic van Hove points of the Brioullin zone (BZ), in agreement with some early theoretical
conjectures \cite{Bok1,SB2}. Additionally, the Fermi surface is unusually flat along one of the
high
symmetry
directions passing through the hyperbolic van Hove point. This
corresponds to the increase of the strenght of the logarithmic van Hove singularity in the electron density of states. Related to it is the early
theoretical proposal of the extended (one-dimensional) van Hove singularities
\cite{Gofron,Abrikosov1,Abrikosov2}, which are stronger than logarithmic.

The experimentally obtained {\bf k}-dependance of the conducting band is usually fitted throughout the BZ with the Fourier series involving only the
first and second harmonics in the symmetry allowed
combinations \cite{Shabel1,Norman1}. Such fits use six independant parameters and they suggest in
particular that the
direct oxygen-oxygen hopping is small, but not negligible with respect to the direct copper-oxygen
hopping of the electrons. The former procedure is unjustly called "tight-binding" fits.

Actually, the proper tight-binding calculation with one copper state and two oxygen states per unit
cell leads to the three band structure which involves the first and higher harmonics in non-linear way. Only when the site energy splitting between
copper and oxygen is by far the largest energy in the problem, the tight-binding structure of the conduction band becomes an additive combination
of low order Fourier harmonics. Here, we will point out that the topology of the measured Fermi surfaces,
expecially those of the weakly dopped LSCO, suggests that the site energy splitting between copper
and oxygen is not large. The analysis of the
electric field gradients in the high-T$_c$ superconductors \cite{Kupcic} led to the same
conclusion. Therefore we carefully examined the tight-binding band
structure characterized with the Cu-O site energy splitting, the Cu-O hopping and the O-O hopping,
for all the ratios of these three parameters. The agreement
with the experiments turns out to be remarkably good, although our fits use
small number of parameters, presumably because the non-linear dependence of the band structure on the first harmonics is properly taken into
account.

The observation of the shadow bands in Bi2212 \cite{Aebi,Ding,Mesot1,Fretwell,Borisenko} opens the
possibility that two of three tight-binding bands are present at the Fermi level. The range
of parameters for which this can occur is determined here. In the three band model this is accompanied by the appearance of the extended van Hove
singularities. It will turn out, however, that the topologies of these two tight-binding Fermi surfaces and the observed Fermi surfaces cannot be
reconciled, suggesting that the shadow band in Bi2212 is the result of spin or of
structure induced doubling of the BZ. This also means that in the
three band model the extended van Hove singularities cannot occur close to the Fermi level, but the usual logarithmic singularities are expected to be strongly
enhanced.

Surprisingly enough, the simple three band tight-binding model \cite{Emery} has not been examined
in sufficient detail before. The previous attempts were focused on the correlation effects
in the high-T$_c$ oxides \cite{Levin1,Levin2,Golosov,Perry1,Perry2}, examining the restricted
range of the tight-binding
parameters. It was concluded in particular that two lower bands can touch
each other in the nighborhood of the Fermi level when the Cu-O site energy splitting and the
oxygen-oxygen hopping are both
positive \cite{Golosov}. Here we point out that such touching can occur for any relative sign of
these parameters and even when the splitting vanishes. Small, doping dependant values of the
renormalized Cu-O
splitting are suggested by our comparison of the three band model with the ARPES
data in the full range of the tight-binding parameters. This emphasizes the need
for the correlation theory which leads to small, doping dependant values of the Cu-O
splitting. It is interesting to note in this respect that the standard ( the O-O
hopping neglected) slave-boson mean field theory does lead to the effective Cu-O
splittings which are small and even change sign for resonable values of the bare tight-binding
parameters \cite{Tutis}. The extensions of those results to the finite O-O hopping is
therefore being carried out and will be presented in separate paper.

\section{General}
\label{I}
\begin{figure}
\begin{center}
\epsfxsize=2in
\epsfbox{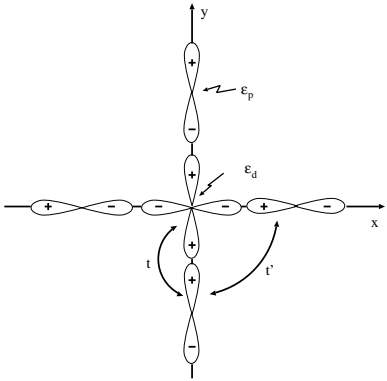}
\end{center}
\caption{Schematic representation of the CuO$_4$ plane defining the band parameters and the 
choice of the phases of the ionic wave functions}
\label{fig1}
\end{figure}

Here we examine the usual, nonmagnetic, undistorted CuO$_2$ square unit cell
with $d_{x^{2}-y^{2}}$ ionic wave function associated to the Cu site, whereas the
oxygen sites on the x and y axes are associated to the $p_x$ and $p_y$ states
respectively. The phases of the wave functions are chosen as in Fig.\ref{fig1}. $\Delta$
is the splitting between oxygen and copper site energies, $\Delta =
\varepsilon_{p}- \varepsilon_{d}$.

\vspace*{0.2cm}
\vspace*{0.2cm}

The hole picture is used here so that $\Delta>0$ corresponds to the
usually quoted values of $\varepsilon_{p}-\varepsilon_{d}$ \cite{Levin1}\cite{Golosov}. The
hybridisations between Cu and O and between
two neighbouring oxygens are introduced through the hopping integrals $t$ and
$t'$
respectively. This simple picture covers the noninteracting holes (electrons) in
the CuO$_2$ plane, but also the case of the very large interaction U$_{d}$ on
the copper site, if the latter is treated by
the mean-field slave boson approximation \cite{Levin1} or by the appropriate Hartree-Fock
\cite{Golosov}. Both approximations lead to the
renormalisation of the bare values of $\Delta$ and $t$, whereas $t'$ is not affected
by correlations.
The Bloch states are built in the usual tight-binding way from $d_{x^{2}-y^{2}}$, $p_x$ and $p_y$ states in each unit cell and the band energies
$\varepsilon_{i}({\bf k})$ are found by diagonalisation of 3x3 hermitean matrix. This
leads to the secular equation of the third order
\begin{equation}
\label{1}
\varepsilon^{3}+3p\varepsilon+2q=0
\end{equation}
where
\begin{equation}
\label{2}
p=-[\alpha+ \beta(\eta^{2}+\xi^{2})+\gamma \eta^{2}\xi^{2}]\:,
\end{equation}
\begin{equation}
\label{3}
q=a+b(\eta^{2}+\xi^{2}) + c\eta^{2}\xi^{2}\:,
\end{equation}
with
\begin{eqnarray}
\label{4}
a&=&
\frac{1}{27} \Delta^{3},\: b=\frac{2}{3} t^{2}\Delta,\:
c=-\frac{16}{3}t'[t'\Delta +3t^{2}], \nonumber\\
\alpha&=&\frac{1}{9}\Delta^{2},\: \beta=\frac{4}{3}t^{2},\:
\gamma=\frac{16}{3}t'^{2}\:,
\end{eqnarray}
and
\begin{equation}
\label{5}
\eta = \sin\frac{k_{x}}{2} \: , \: \xi =\sin\frac{k_{y}}{2}\:.
\end{equation}
$\varepsilon $ is the energy measured with respect to $\widetilde\varepsilon
=(\varepsilon_{d}+2\varepsilon_{p})/3$.
The hermiticity of 3x3 tight-binding matrix ensures that three roots
$\varepsilon_{i}$ of Eq.(\ref{1}) are
real. It should be noted that Eq.(\ref{1}) is invariant on the simultaneus
replacement $\Delta \rightarrow -\Delta , t'\rightarrow -t'$ and $\varepsilon \rightarrow
-\varepsilon$, irrespectively of the sign of $t$. This means that the band structure
$\varepsilon_{i}({\bf k})$ for $\Delta<0,t'<0 $ or for $\Delta>0, t'<0$ can be obtained
from the $\varepsilon_{i}({\bf k})$ for $\Delta >0, t'>0 $ or for $\Delta <0,
t'>0$ respectively by the reflection on the {\bf k} plane. We shall therefore, without
the loss of generality, carry out the theoretical discussion for $t,t' >0$ and
for both signs of $\Delta$, although $t'<0$ will also suit the
experimental situation, as discussed in Section \ref{III}. It can be finally noted that the band structure in
the electron language, appropriate for the ARPES analysis is obtained from the band
structure in the hole language used here by the reflection on the {\bf k} plane.

\section{Small $t'$}
\label{II}
\begin{figure}
\begin{center}
\epsfxsize=2in
\epsfbox{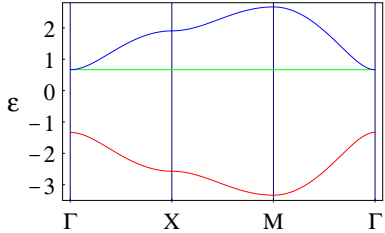}
\end{center}
\caption{Usual band structure when $t'=0,$ $\Delta >0$, energy in units $t$}
\label{fig2}
\end{figure}

It is instructive to start the discussion of the features of the energy bands by introducing t' as a small quantity.
The well known results for $t'=0$ are shown in Fig.\ref{fig2}. For $\Delta >0$ lower band
$\varepsilon_{L}$ is splitted from the intermediate and the upper band
$\varepsilon_{I},\varepsilon_{U}$ by $\Delta$ and the
latter two bands, $\varepsilon_{I}$ being dispersionless, are degenerate at the ${\bf
\Gamma}$ point. For $\Delta <0$, the dispersionless band joins the lower band
coinciding with it at the ${\bf \Gamma}$ point. The lowest band $\varepsilon_{L}$ is half
filled in the stoechiometric crystal, and the Fermi energy $\varepsilon{^{0}_{F}}$
coincides with the van Hove singularities at $\varepsilon{^{L}_{x}}$, for any value of t
and $\Delta$.

\vspace*{0.2cm}
\vspace{0.2cm}

\begin{figure}
\begin{center}
\epsfxsize=2in
\epsfbox{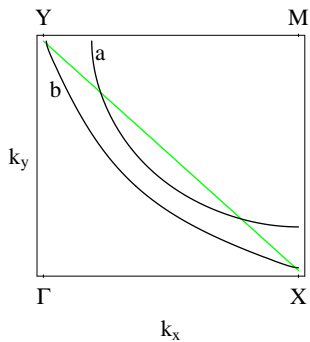}
\end{center}
\caption{Equienergetic contours of $\varepsilon_{L}$ obtained for $\Delta <0$ and small $t'\geq
0$; green
- equienergetic line at $t'=0$ that passes through van Hove points at
$\varepsilon{_{x}^{L}}$, (a) - energy $\varepsilon^{0}$ that separates the states below and above it in two halves,
(b) - equienergetic curve that passes through van Hove points} 
\label{fig3}
\end{figure}

The small parameter $t'$ can be introduced either by the usual perturbation theory
applied to 3x3 tight-binding matrix or by iteration of the solutions of
Eq.(\ref{1}). Retaining only the leading term, linear in $t'$, (i.e. the first order perturbation
theory) gives the structure of $\varepsilon_{L}$
\begin{equation}
\label{6}
\varepsilon_{L}({\bf k})=\varepsilon{^{0}_{L}}({\bf k})+\frac {32t't^{2}\eta ^{2}\xi
^{2}}{A(2A+\varepsilon_{p}+\varepsilon_{d} )}\:,
\end{equation}
where
\begin{equation}
2A=\sqrt{\Delta^{2} + 16t^{2}(\eta^{2}+\xi^{2})}\:.
\end{equation}
The main qualitative effect of $t'$ is the change in the number of the states below
the van Hove singularity at $\varepsilon{^{L}_{x}}$, the value of which remains itself
unaffected by $t'$. For $\Delta <0$ or $\Delta >0$ the number of the states below
$\varepsilon{^{L}_{x}}$
is increased or decreased respectively. In both cases the Fermi level $\varepsilon{^{0}_{F}}$
of the half-filled band is removed from the van Hove singularity at 
$\varepsilon{^{L}_{x}}$. The quantitative discussion is based on Fig.\ref{fig3}.

\vspace*{0.2cm}
\vspace*{0.2cm}

For$\Delta <0$ the $\varepsilon{^{L}_{x}}$ equienergetic straight line is bent towards the ${\bf \Gamma}$ point
and all the states between this line and the finite $t'$ curve are transferred from below  $\varepsilon{^{L}_{x}}$ to the energies above it. The number of the
transferred states can be determined for small $t'$ by expanding in terms of the
small departure of the $\varepsilon{^{L}_{x}}$ equienergetic curve for $t'\neq 0$ from
the straight line and by calculating the surface between the straight line and this curve. Thus
obtained number of the transferred states $\delta_{c}$ is
\begin{mathletters}
\label{7}
\begin{equation}
\label{7a}
\delta_{c}=-\frac{4t'}{\pi^{2}t} {\rm sign} \Delta,\:\:\:\:\:
t'<|\Delta| <t,
\end{equation}
and
\begin{equation}
\label{7b}
\delta_{c}=\frac{-8t'}{\pi^{2}\Delta },\:\:\:\:\:
t'|\Delta| <t^{2} < \Delta^{2}.
\end{equation}
\end{mathletters}
The energy $\varepsilon^{0}$ which divides the number of the states in the
band in two halves can be calculated in the same spirit
by finding the equienergetic curve which encompasses the same number of states
below and above the straight line, as shown in Fig.\ref{fig3}. For the two regimes
considered above
\begin{mathletters}
\label{8}
\begin{equation}
\label{8a}
(\varepsilon^{0}-\varepsilon{^{L}_{x}})\ln\frac{|\varepsilon^{0}-\varepsilon{^{L}_{x}}|}{4t}=2t' 
{\rm sign} \Delta \:,
\end{equation}
 and
\begin{equation}
\label{8b}
\frac{\Delta
(\varepsilon^{0}-\varepsilon{^{L}_{x}})}{t^{2}}\ln\frac{\Delta
(\varepsilon^{0}-\varepsilon{^{L}_{x}})}{t^{2}}=
\frac{16t'}{\Delta}\:.
\end{equation}
\end{mathletters}
Note that $\varepsilon^{0}$ coincides with the Fermi level when the band $\varepsilon_{L}$ is half
filled, and then $\delta_{c}$ is the critical hole doping which brings the Fermi level to
$\varepsilon{^{L}_{x}}$.

The number of the states $\delta$ between the two close energies $\varepsilon$ and
$\varepsilon^{0}$ can be determined by similar reasoning as
\begin{mathletters}
\label{9}
\begin{equation}
\label{9a}
\varepsilon-\varepsilon^{0}= \frac{\pi^{2}t \delta
}{2\ln\frac{|\varepsilon^{0}-\varepsilon{^{L}_{x}}|}{4t}}{\rm sign} \Delta\:,
\end{equation}
and
\begin{equation}
\label{9b}
\varepsilon-\varepsilon^{0}=\frac{\pi^{2}t^{2}\delta}{8\Delta
\ln\frac{\Delta(\varepsilon^{0}-\varepsilon{^{L}_{x}})}{t^{2}}}\:.
\end{equation}
\end{mathletters}
Obviously, Eq.(\ref{9}) also defines the position of the Fermi level at
$\varepsilon=\varepsilon_{F}$, if $\delta$
is taken to be the doping of the CuO$_2$ plane.

In two considered limis Eqs.(\ref{9}) can be used to determine the density of states
$d\delta/dE$ for $\varepsilon$ between $\varepsilon{^{L}_{x}}$ and
$\varepsilon^{0}$
\begin{mathletters}
\label{10}
\begin{equation}
\label{10a}
n_{L}(\varepsilon)=\frac{2}{\pi^{2}t}(1+\frac{2t'}{t})\ln\frac{|\varepsilon
-\varepsilon{^{L}_{x}}|}{4t}\:,
\end{equation}
and
\begin{equation}
\label{10b}
n_{L}(\varepsilon)=\frac{|\Delta|}{2\pi^{2}t^{2}}(1+\frac{4t'}{|\Delta|})\ln\frac{\Delta
(\varepsilon
-\varepsilon{^{L}_{x}})}{t^{2}}\:,
\end{equation}
\end{mathletters}
or vice versa, Eqs.(\ref{10}) and (\ref{8}) can be used to derive
Eq.(\ref{9}). The density of states between $\varepsilon^{0}$ and
$\varepsilon{^{L}_{x}}$ is increased by $t'$.

It should be noted that the results (\ref{8}) to (\ref{10}) concern only the energetic
neighbourhood of the van Hove energy
$\varepsilon{^{L}_{x}}$. The behavior of the band structure far from $\varepsilon{^{L}_{x}}$ is
irrelevant in this respect, in spite of the fact that t' leads to the level
crossing for all the energies down to $\varepsilon{^{L}_{M}}$.

Let us close this discussion by noting that the density of states can be obtained from Eq.(\ref{6}) in the
closed form for all energies when $\Delta >t',t$. In this case, the spectrum of
the lowest band can be obtained explicitly,
\begin{mathletters}
\label{11}
\begin{equation}
\label{11a}
\varepsilon_{L}({\bf
k})=-\frac{2}{3}\Delta +\frac{4t^{2}}{\Delta}(\eta^{2}+\xi^{2}-\frac{8t'}{\Delta}
\eta^{2}\xi^{2})
\end{equation}
in agreement with the usual additive form of the "tight-binding" fits. This leads to the
density of states \cite{Xing}
\begin{equation}
\label{11b}
n_{L}(\varepsilon)=\frac{2|\Delta|}{\pi^{2}t}\frac{1}{\sqrt{t^{2}-32t'\varepsilon}}{\bf
K}[\frac{8\varepsilon(t^{2}|\Delta|-4t't^{2}-2\varepsilon\Delta^{2})}{t^{3}\sqrt{t^{2}-32t'\varepsilon}}]
\end{equation}
\end{mathletters}
where the first term in Eq. (\ref{11a}) is omitted and $K(x)$ is the complete elliptic integral of
the first kind.
For $t'|\Delta|<t^{2}<\Delta^{2}$, Eq.(\ref{11b}) reproduces the result (\ref{10b}).

\section{Band degeneracy}
The main aim of the discussion which follows is to consider the degeneracies
or the near degeneracies of the three bands $\varepsilon_{i}$, at a given {\bf
k}, which are caused by including the oxygen-oxygen hopping $t'$ into the
dispersions in non-perturbative way. It is clear from Fig.\ref{fig3}
that such effects can occur only when the dispersions related to $t$ and $t'$ are at
least of the same order of magnitude, i.e. for the sufficiently large $t'$.

The large $t'$ is also related to the qualitative modification of the dispersion
within each band itself, e.g. the bending of the $\varepsilon{^{L}_{x}}$
equienergetic line, instead of being small, as in Fig.\ref{fig3}, becomes large, obviously
changing the symmetry properties of the equienergetic surfaces all over the BZ.
This is related to the fact that the O-O overlap induces the hopping along
the diagonal of the Cu-O unit cell, unlike the Cu-O overlap, associated with the
propagation along the cell's main axes, i.e. the additional propagation axes
are rotated by $\pi/4$ with respect to the original ones.

As well known, the quantity proper to the study of the degeneracy of the spectrum
$\varepsilon_{i}$ at a given {\bf k} is
\begin{equation}
\label{12}
D=q^{2}+p^{3},
\end{equation}
known as the discriminant of the third order Eq.(\ref{1}). The hermiticity of the
problem, i.e. of 3x3 tight-binding matrix, which makes the roots $\varepsilon_{i}$
real, ensures that
\begin{equation}
\label{13}
D \le 0
\end{equation}
For $D<0$, all the roots $\varepsilon_{i}$ are real and different, whereas $D=0$,
with $p^{3}=-q^{2}\neq 0$ implies that two roots are degenerate. For $p=q=0$, all three
roots coincide. Together with the inequality (\ref{13}), this means that the zero
of $D$ at a given point {\bf k} of the BZ is also the maximum of $D$. Therefore,
instead of the direct calculation of the zeroes of $D$, which is the third order
polinomial in $\eta^{2}$ and $\xi^{2}$, it is possible to find the optima of $D$ with
respect to $\eta$ and $\xi$, and to determine which of those, if any, are the
zeros of $D$. In carrying out this procedure, it is convenient to use the polar
coordinates instead of $\eta$ and $\xi$ of Eq.(\ref{5})
\begin{equation}
\label{14}
\eta =\rho\sin\varphi \: , \: \xi =\rho\cos\varphi\:.
\end{equation}
The zero of the derivative of D with respect to $\varphi$ is easily found to be
\begin{eqnarray}
\label{15}
\frac{\partial D}{\partial
\varphi}&=&\rho^{4}\sin
4\varphi[c(a+b\rho^{2}+\frac{c\rho^{4}\sin^{2}2\varphi}{4})
\nonumber \\
&-&\frac{3\gamma}{2}(\alpha
+\beta \rho^{2}+\frac{\gamma \rho^{4}\sin^{2}2\varphi}{4})^{2}]=0.
\end{eqnarray}
The advantage of this representation is that, besides separating out immediately the
trivial solution at $\rho=0$, which corresponds to the degeneracy at the ${\bf
\Gamma}$ point, it also exibits other possible band degeneracies at
$\varphi_{1,2}=0, \pi/4$, i.e. on the ${\bf \Gamma X}$ and ${\bf \Gamma M}$ axes of the BZ.

\begin{figure}
\begin{center}
\epsfxsize=2in
\epsfbox{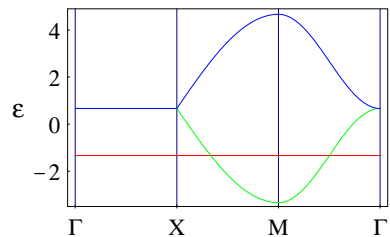}
\end{center}
\caption{Energy structure bands when $t=0$, separated in two oxygen bands and the 
dispersionless copper
level, energy in units $t'$}
\label{fig4}
\end{figure}

The nature of the third maximum, associated with the zero of the square braces in
Eq.(\ref{15}) is best understood for $t=0$, when this zero occurs for
\begin{equation}
\label{16}
\rho^{2}\sin2\varphi =\frac{|\Delta|}{2t'}\:.
\end{equation}
This is easily recognized as the condition that the pure oxygen band crosses the
dispersionless copper level, as shown in Fig.\ref{fig4}. Consequently, $D_{0}=D(t=0)=0$ on the
line (\ref{16}).

\vspace*{0.2cm}
\vspace*{0.2cm}

\begin{figure}
\begin{center}
\epsfxsize=1.5in
\epsfbox{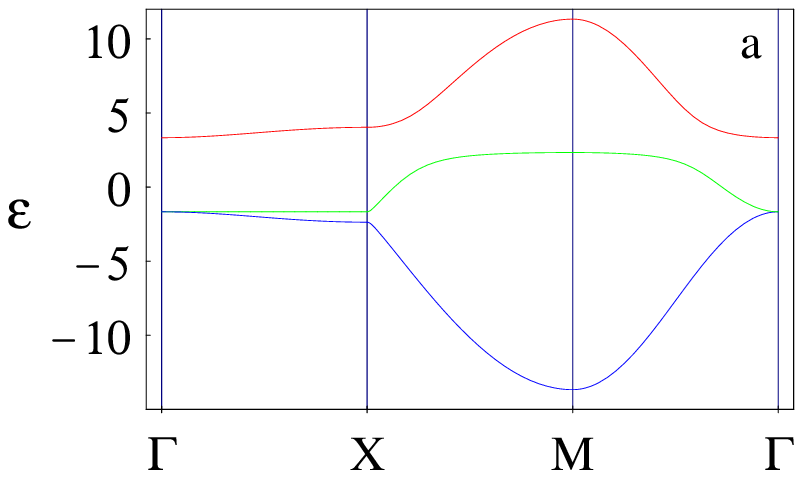}
\end{center}
\begin{figure}
\begin{center}
\epsfxsize=1.5in
\epsfbox{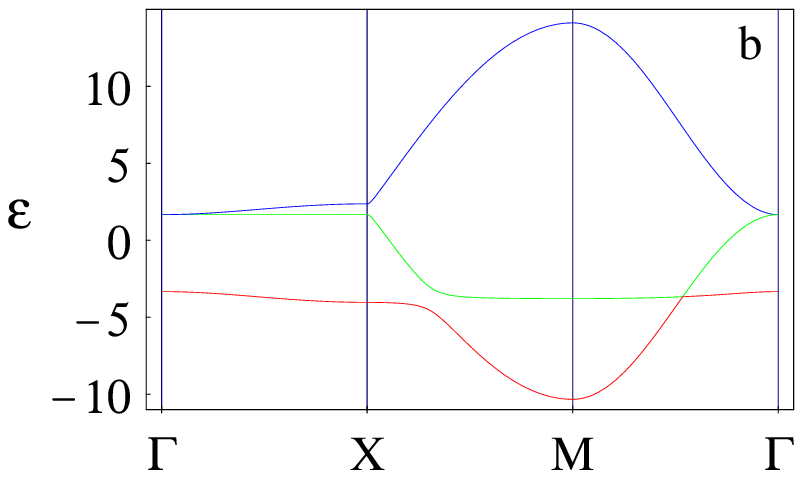}
\end{center}
\caption{Band anticrossing  when
$t'/t=3$ and $\Delta/t$=-5 (a) and touching for $t'/t=3$ and $\Delta/t$=5 (b),
energy in units $t$}
\label{fig5}
\end{figure}
\end{figure}

In the next step the change $\Delta D$ is calculated to the leading order in $t$. On the
line (\ref{16})
\begin{equation}
\label{17}
\Delta D=-\frac{32}{27}\Delta^{4}t^{2}(\rho^{2}-\frac{\Delta}{2t'})\:.
\end{equation}
It follows that $\Delta D<0$ in all the points on the crossing line (\ref{16})
except for $\Delta >0$ in the single point on the ${\bf\Gamma M}$ diagonal
\begin{equation}
\label{18}
\varphi_{3} =\frac{\pi}{4},\qquad \rho{^{2}_{3}}=\frac{\Delta}{2t'}\:,
\end{equation}
where $\Delta D=0$, i.e. $D=D_{0}+\Delta D=0$. The band degeneracy associated
with the third maximum of D occurs on the ${\bf\Gamma M}$ diagonal and it is possible only for the same sign of
$\Delta$ and $t'$. The variation of $\Delta D$ in the vicinity of the line
(\ref{16}) is weak with respect to the variation of
$D_{0}<0$, i.e. $D=D_{0}+\Delta D<0$ in the vicinity of the line (\ref{16}). Therefore,
small $t$ removes the degeneracy of two bands, except at the point (\ref{18}), rather
than displacing slightly the crossing line (\ref{16}) to the new position. In other
words, small $t$ introduces the anticrossing of the oxygen band and the former copper level,
excepting the single point (\ref{18}) for $\Delta >0$, as exemplified in Fig.\ref{fig5}a
and \ref{fig5}b.

\vspace*{0.2cm}
\begin{multicols}{2}
\end{multicols}
\vspace*{0.2cm}

Turning back to the first, $\varphi_{1}=0$ maximum of $D$, on the ${\bf \Gamma X}$ line, we note
that for $t=0$ it corresponds to the degeneracy of two oxygen bands, symmetric by
reflection on the {\bf k} plane, while the dispersionless copper level is decoupled from
the oxygen bands. When the copper is removed, the natural BZ corresponds to one oxygen per unit
cell, i.e. it is rotated for $\varphi/4$ with respect to the original
one and increased by $\sqrt{2}$. In this zone two bands unfold into one continuous
oxygen band and the dispersionless ${\bf \Gamma X}$ line becomes analoguous to the
dispersionless straight line in Fig.\ref{fig3} for $t=0$. The corresponding singularity in the
density of
states is therefore logarithmic \cite{Bok1,Bok2}. For $t\neq 0$, the degeneracy on the ${\bf \Gamma X}$
line is removed, except in the ${\bf \Gamma}$ point, as shown e.g. in Fig.\ref{fig5}b, with the
consequences further
discussed in Sec.\ref{III}

\subsection{Lower critical value of t'}

\begin{figure}
\begin{center}
\epsfxsize=2in
\epsfbox{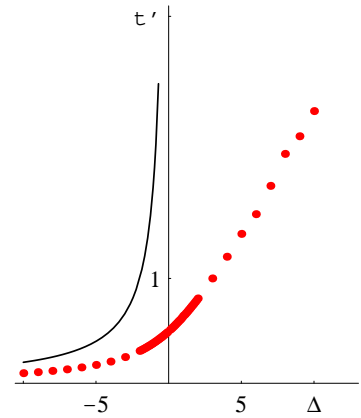}
\end{center}
\caption{For $t'$ increasing from the lower critical value $t'_{cr}$ to the upper critical value
$\widetilde{t'_{cr}}$ the band touching that starts at the {\bf M} point moves along the ${\bf
\Gamma
M}$ line towards
the ${\bf \Gamma}$ point. Limiting behaviors for small and large $\Delta/t$ described in the
text are completed numerically (red dots). $\Delta$ and $t'$ in units $t$.}
\label{fig6}
\end{figure}

The previous discussion shows that for finite $t$,$t'$ the degeneracies of the bands at a given {\bf k}
point can occur only on the ${\bf \Gamma M}$ line. Setting $\eta = \xi$ in the equation $D(\eta ,\xi )=0$ simplifies
considerably further discussion. To this effect, it is convenient to distinguish between
$\Delta >0$ and $\Delta <0$, as already suggested by Eq.(\ref{17}) and start with the limit
of small $|\Delta |$. The solution of the equation $D=0$ is particulary simple
around $\Delta =0$
at the {\bf M} point of the BZ. By increasing $t'$ from zero, the dispersionless oxygen
level $\varepsilon_{I}$ of Fig.\ref{fig2} bends towards the lower band $\varepsilon_{L}$ and
these two bands touch at the {\bf M} point. It can be
shown analitically that, for $|\Delta |$ small, the degeneracy at the {\bf M} point occurs for
\begin{equation}
\label{19}
\frac{t'_{cr}}{t}=0.5+0.167\frac{\Delta}{t}\:,
\end{equation}
i.e. for $\Delta >0$ the oxygen level must bend more to cross the gap $\Delta $ than for
$\Delta <0$, when there is no gap to cross. Eq.(\ref{19}) shows that the band
touching occurs not only for $\Delta >0$, as already known \cite{Golosov}, but also for $\Delta
<0$. This is allowed by the solution $\varphi_{2}=\pi/4$ of the
Eq.(\ref{15}), although the solutions $\varphi_{3}=\pi/4,\rho_{3}$,
given by Eq.(\ref{18}), is not a zero of $D$ when $\Delta <0$. The full curve
$t'_{cr}/t$ versus $\Delta /t$ is
obtained numericaly in Fig.\ref{fig6}. The limit of the large $\Delta /t$ in
Fig.\ref{fig7} is
transparent:
it corresponds to the situation when the pure oxygen band touches the flat copper level at
the {\bf M} point of the BZ. According to the Eq.(\ref{16}), this occurs for
$4t'_{cr}=\Delta $. In Fig.\ref{fig7}a and \ref{fig8}a, the full three band structures,
obtained
by
solving
Eq.(\ref{1}) numerically, are displayed for $\Delta /2t =\pm 0.5$ when
$t'/t=t'_{cr}/2t=0.64$ and 0.39 respectively.

\vspace*{0.2cm}
\vspace*{0.2cm}

\subsection{Upper critical value of t'}

\begin{figure}
\begin{center}
\epsfxsize=1.5in
\epsfbox{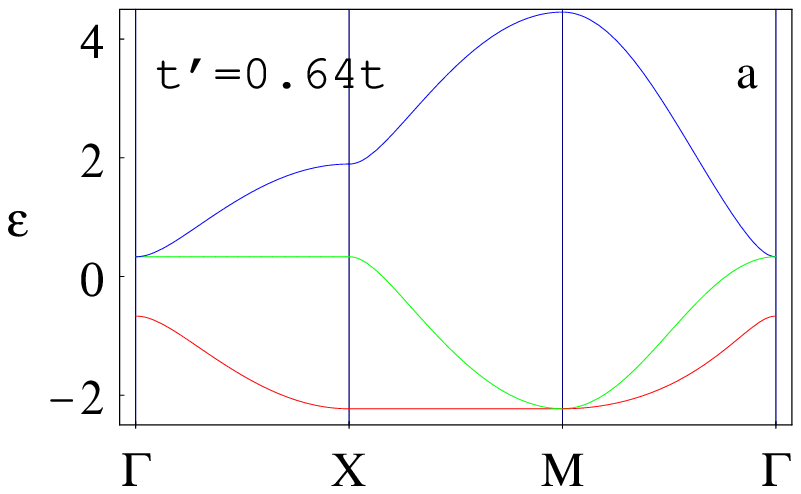}
\end{center}
\begin{figure}
\begin{center}
\epsfxsize=1.5in
\epsfbox{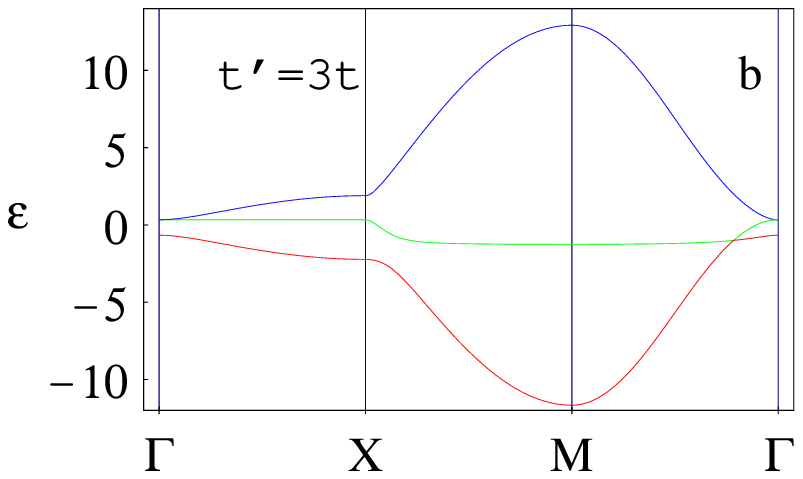}
\end{center}
\caption{Illustration of the touching of the two lower bands when $t'$and $\Delta$ are of the same
sign for $\Delta/t=1$. The touching
point moves from {\bf M} point (a) towards ${\bf \Gamma}$
point (b), when the value of $t'$ is increased from its critical value $t'_{cr}$.
Energy in units $t$.}
\label{fig7}
\end{figure}
\end{figure}

It is further interesting to investigate what happens for $t'>t'_{cr}$ for the fixed value of
$\Delta/t >0 $ in Fig.\ref{fig6}. Some representative band structures are shown in
Fig.\ref{fig7} for $\Delta
/t=1$. It appears that the bands $\varepsilon_{I}$ and $\varepsilon_{L}$ touch at a point
which moves from the {\bf M} point towards the ${\bf \Gamma }$ point as t' increases from
$t'_{cr}$. For $t'/t \gg 1$, the regime of the Eq.(\ref{16}),(\ref{17}) and
(\ref{18}) is reached, i.e. the touching of the bands occurs at $\rho^{2}
=\Delta /2t' <1$.

\vspace*{0.2cm}
\begin{multicols}{2}
\end{multicols}
\vspace*{0.2cm}

\begin{figure}
\begin{center}
\epsfxsize=1.5in
\epsfbox{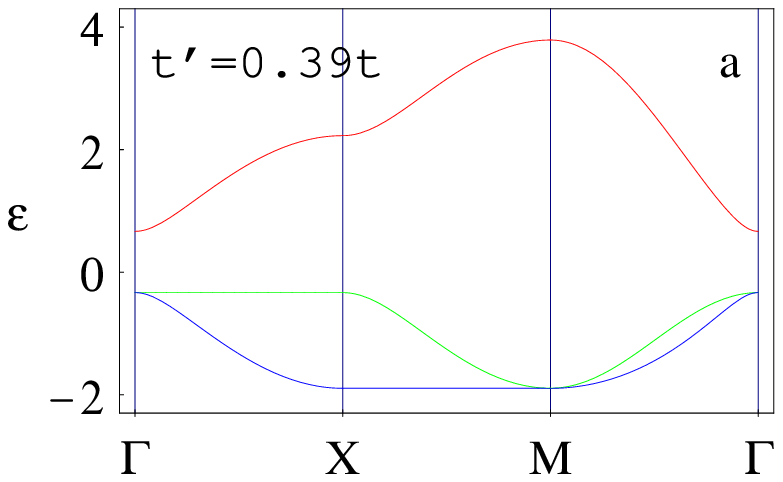}
\end{center}
\begin{figure}
\begin{center}
\epsfxsize=1.5in
\epsfbox{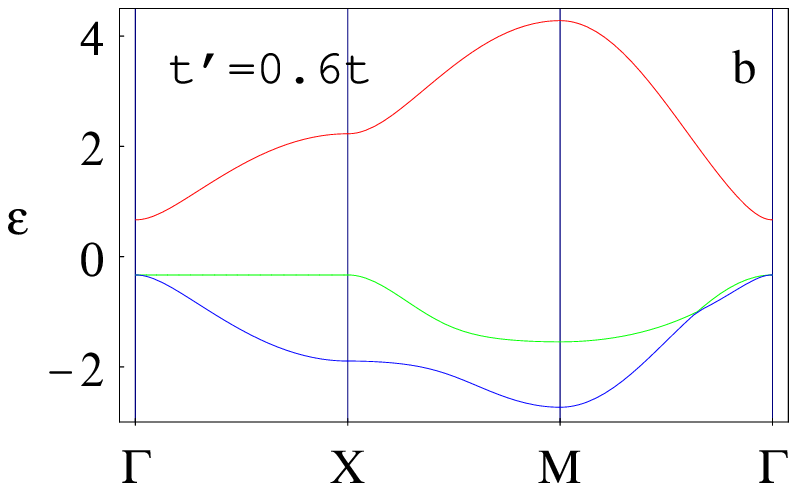}
\end{center}
\begin{figure}
\begin{center}
\epsfxsize=1.5in
\epsfbox{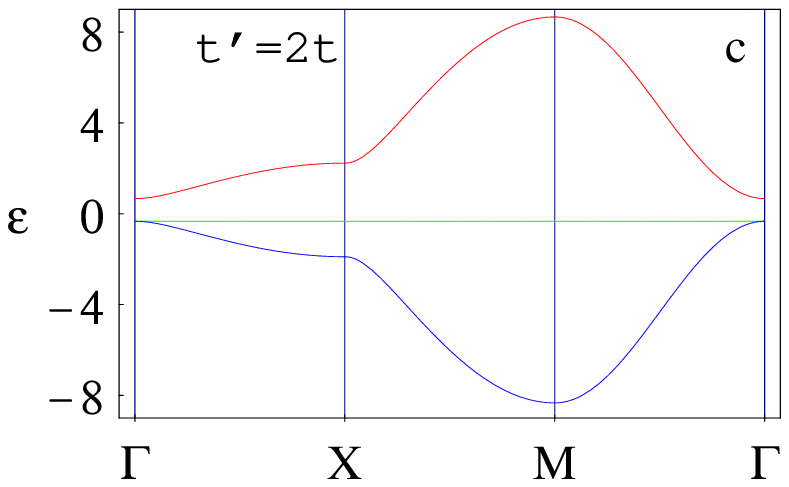}
\end{center}
\begin{figure}
\begin{center}
\epsfxsize=1.5in
\epsfbox{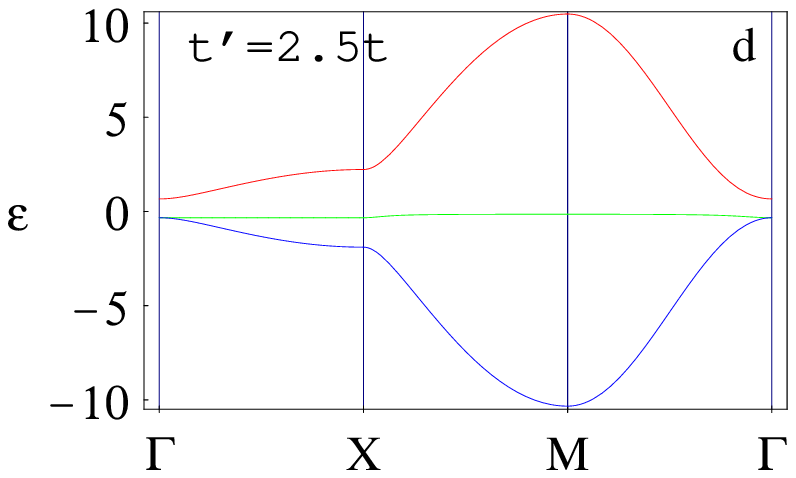}
\end{center}
\caption{Illustration of the touching of the two lower bands when $t'$ and $\Delta$ are of the
opposite sign for $\Delta/t=-1$. The
touching point of the two lower bands moves from {\bf M} point (a) towards ${\bf
\Gamma}$ point when the value of $t'$ is increased from $t'_{cr}$ to
$\widetilde{t'_{cr}}$ at which the intermediate band becomes flat (c). With further increment of
$t'$ this band bends upwards (d). Energy in units $t$.}
\label{fig8}
\end{figure}
\end{figure}
\end{figure}
\end{figure}

The situation is more complicated for $\Delta <0$. It is illustrated in Fig.\ref{fig8} for
$\Delta/t =-1$. As
$t'/t$ increases beyond $t'_{cr}/t$, the point of the coincidence moves towards the ${\bf \Gamma }$
point, reaching and leaving it for
\begin{equation}
\label{20}
\frac{\widetilde{t'_{cr}}}{t} =-\frac{2t}{\Delta }.
\end{equation}
For this value of parameters $\varepsilon _{I}$ is again dispersionless as it
was before for $t=0$ in Fig.\ref{fig4}. The corresponding dispersions of the bands $\varepsilon
_{L}$
and
$\varepsilon _{U}$ are given by
\begin{equation}
\label{21}
\varepsilon _{L,U}=-\frac{\Delta}{2}[1\pm
\sqrt{1+\frac{16t^{2}}{\Delta^{2}}(\eta^{2}+\xi^{2}+16\eta^{2}\xi^{2})}\: ]\: .
\end{equation}
For $t'>\widetilde{t'_{cr}}$, $\varepsilon _{I}$ starts to bend towards $\varepsilon _{U}$ and,
according to the Eq.(\ref{18}), reaches the anticrossing with this band at $\rho
^{2}$ given approximately by Eq.(\ref{16}). This anticrossing is more pronounced
for $\Delta/2t =-5$ of Fig.\ref{fig5}b than for $\Delta/2t=-1$ of
Fig.\ref{fig8}d.

\vspace*{0.2cm}
\begin{multicols}{2}
\end{multicols}
\vspace*{0.2cm}
 
\begin{figure}
\begin{center}
\epsfxsize=1.5in
\epsfbox{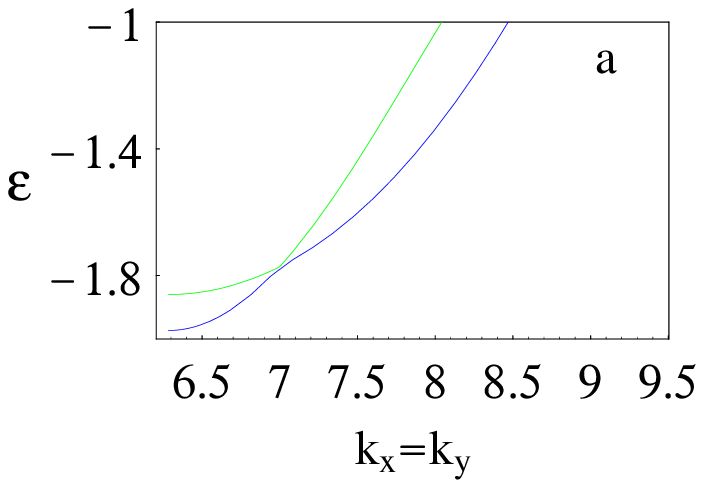}
\end{center}
\begin{figure}
\begin{center}
\epsfxsize=1.5in
\epsfbox{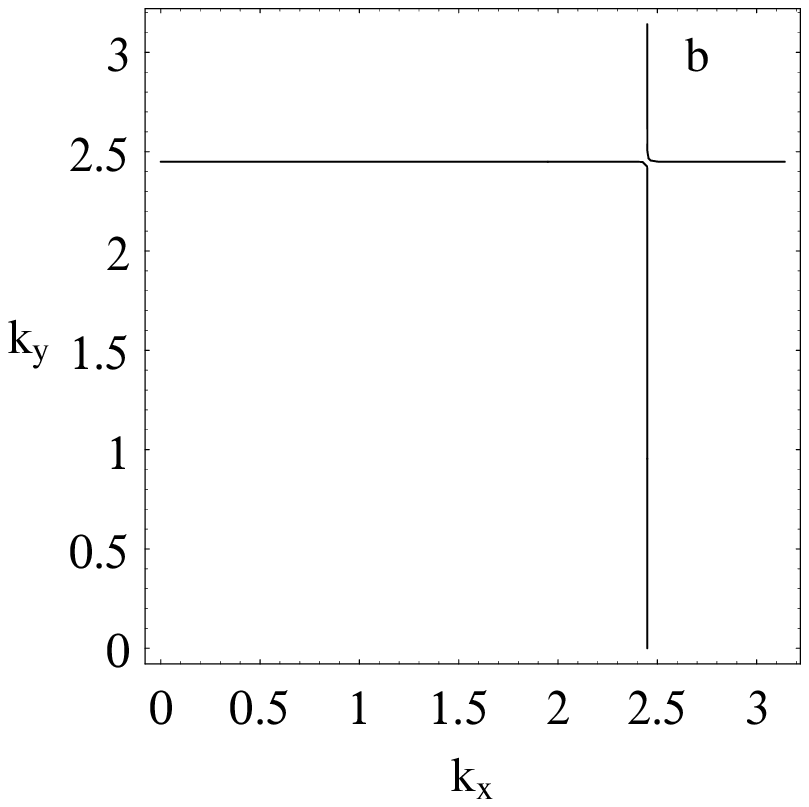}
\end{center}
\begin{figure}
\begin{center}
\epsfxsize=1.5in
\epsfbox{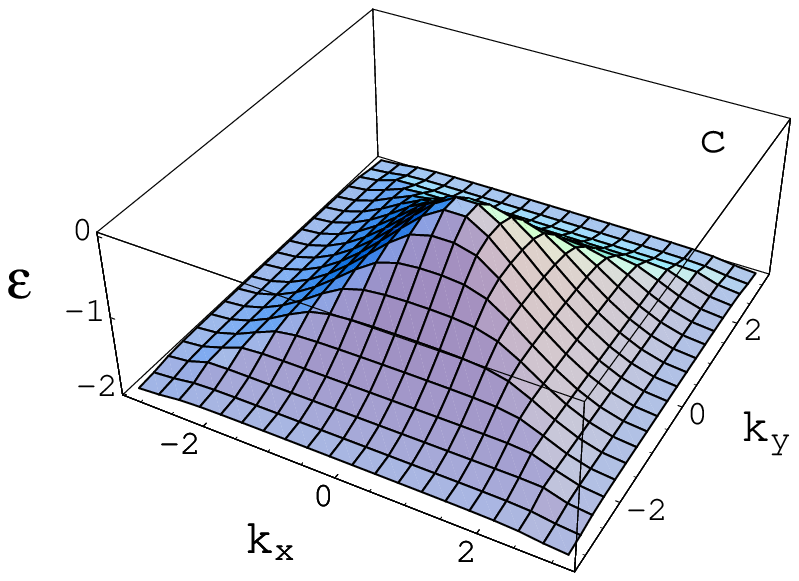}
\end{center}
\begin{figure}
\begin{center}
\epsfxsize=1.5in
\epsfbox{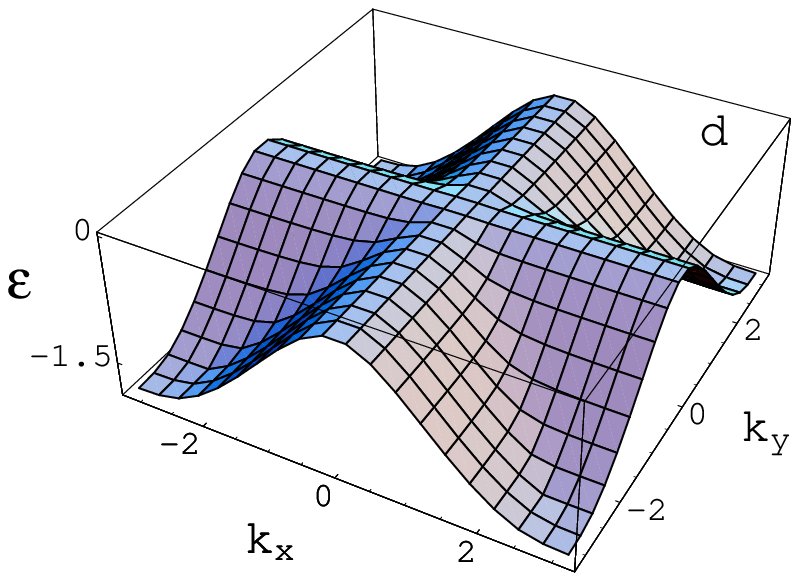}
\end{center}
\caption{The band touching of Fig.8d with better resolution (a), the equienergetic surfaces of the
bands $\varepsilon_{L}$¢and $\varepsilon _{I}$ when they touch (b) and the full
dispersion of the bands $\varepsilon _{L}$ and $\varepsilon _{I}$ (c and d).
Energy in units $t$. }
\label{fig9}
\end{figure}
\end{figure}
\end{figure}
\end{figure}

The discussion is completed by showing in Fig.\ref{fig9}a, the band touching of
Fig.\ref{fig8}b
with better resolution, the equienergetic surfaces of the bands
$\varepsilon_{L}$¢and $\varepsilon _{I}$ when they touch in the Fig.\ref{fig9}b and the full
dispersion of the bands $\varepsilon _{L}$ and $\varepsilon _{I}$ in Figs.\ref{fig9}c and d. The
topology of the bands on Fig.\ref{fig9}b for $\Delta<0$ shows striking similarity to the one found
before for the bands that touch at $\Delta >0$, i.e. the observation of such
topology cannot be used to select the sign of $\Delta $.

\vspace*{0.2cm}
\begin{multicols}{2}
\end{multicols}
\vspace*{0.2cm}

\subsection{$\Delta =0$ bands}
The features described above for $\Delta <0$ and $\Delta >0$ can be studied
explicitely on the borderline $\Delta=0$ between those two (cf. Fig.\ref{fig6}) because
then the electron spectrum can be obtained in the closed form
\begin{mathletters}
\label{22}
\begin{equation}
\label{22a}
\varepsilon_{L}=-4t'\sqrt{\eta^{2}\xi^{2}+\frac{t^{2}}{4t'^{2}}(\eta^{2}+\xi^{2})}+\frac{t^{2}}{t'}\frac{\eta^{2}\xi^{2}}{\eta^{2}\xi^{2}+\frac{t^{2}}{4t'^{2}}(\eta^{2}+\xi^{2})}\:,
\end{equation}
\begin{equation}
\label{22b}
\varepsilon_{I}=\frac{-2t^{2}}{t'}\frac{\eta^{2}\xi^{2}}{\eta^{2}\xi^{2}+\frac{t^{2}}{4t'^{2}}(\eta^{2}+\xi^{2})}\:,
\end{equation}
\begin{equation}
\label{22c}
\varepsilon_{U}=4t'\sqrt{\eta^{2}\xi^{2}+\frac{t^{2}}{4t'^{2}}(\eta^{2}+\xi^{2})}+\frac{t^{2}}{t'}\frac{\eta^{2}\xi^{2}}{\eta^{2}\xi^{2}+\frac{t^{2}}{4t'^{2}}(\eta^{2}+\xi^{2})}\:.
\end{equation}
\end{mathletters}
for $t'>>t$ and similarly for $t'<<t$.

Eqs.(\ref{6}),(\ref{21}) and (\ref{22}) illustrate the fact that in the tight-binding approximation the spectrum is built from low order harmonics
$\eta^{2},$ $\xi^{2}$ of the Eq.(\ref{5}) in nonlinear way. Eq.(\ref{22}) and (\ref{23}) are
useful also because it will turn out in Sec.\ref{IV} that the experimental specra
are close to the $\Delta =0$ limit.

\section{Extended van Hove singularities}
\label{III}

Inspection of Figs.\ref{fig7} and \ref{fig8} shows that the degeneracy of the bands
$\varepsilon
_{L}$ and
$\varepsilon _{I}$ at the point {\bf M}, for $t'=t'_{cr}$ is acompanied in the band
$\varepsilon _{L}$ with the absence of the dispersion on the line
between ${\bf \Gamma}$ and {\bf M} points. It appears that the dispersion depends only on the component of the wavevector perpendicular
to the ${\bf \Gamma M}$ line, in a quadratic way, i.e. the dispersion is one-dimensional on
whole ${\bf \Gamma M}$ line. As a result, an "extended" van Hove singularity is found in
the density of states\cite{Gofron,Abrikosov1,Abrikosov2} of the lowest band
\begin{equation}
\label{23}
n_{L}(\varepsilon ) \approx \frac{B}{\sqrt{\varepsilon -\varepsilon
{^{M}_{L}}}}.
\end{equation}
When the degeneracy between the ${\bf \Gamma }$ and {\bf M} points is lifted by changing
slightly $t'$ from $t'_{cr}$,  the usual step singularity at $\varepsilon {^{L}_{M}}$ and the
logarithmic singularity at $\varepsilon {^{L}_{x}}$ are recovered in
$n_{L}(\varepsilon )$, lying close to each other in energy, i.e. enhancing each other. The
described coalescence
of these two singularities can be explicitly found in Eq.(\ref{11}), but it occurs there for
$-8t'=\Delta $, when, unfortunately, the expansion in terms of $|\Delta | \gg t,t'$ is only
qualitatively correct.

Similar extended van Hove singularities are associated with the absence of the dispersion on the
${\bf \Gamma X}$ line in the band $\varepsilon _{I}$ for any finite values of
$t$ and $t'$. For $t=0$
however, as already explained below Eq.(\ref{17}), the copper level is decoupled from the
oxygen band structure and two oxygen bands $\varepsilon_{I}$ and $\varepsilon_{U}$ coalesce
into one band with the artificial
degeneracy on the ${\bf \Gamma X}$ line. When the artificial degeneracy is removed, by
increasing the BZ so that it corresponds to one oxygen per unit cell, the spectrum of this band is
continuously increasing through the ${\bf \Gamma X}$ line, what is thus associated with the
logarithmic, and not with the extended van Hove singularity in the density of states.

It is apparent here that in the three band model with $1+ \delta $ holes, the Fermi level
falls far from the extended van Hove singularity in the $\varepsilon
_{I}$ band. On the other hand, when the extended van Hove
singularity (\ref{23}) occurs for $t'=t'_{cr}$ in the $\varepsilon _{L}$ band, the Fermi level
crosses necessarily two bands, $\varepsilon _{L}$ and $\varepsilon _{I}$. This fact makes the
discussion of the main band and the shadow bands, observed in some of the high-$T_{c}$
superconductors \cite{Aebi,Ding,Mesot1,Fretwell,Borisenko},
quite relevant. The rest of this paper is devoted to the analysis of the empirical
band structure. As already mentioned, it will turn out that the observed
dispersions of the main band and the shadow band cannot be reconciliated with
$\varepsilon _{L}({\bf k})$ and
$\varepsilon _{I}({\bf k})$, i.e. the extended van Hove singularities of the three band
model do not occur close to the Fermi level.

\section{Experiments}
\label{IV}
The previous discussion provided the complete description of the three band structure. Although it was carried out for $t'>0$ and arbitrary sign of
$\Delta$, the symmetry properties (cf Sec.\ref{I}) extend it to the entire range of the parameters. In
this section, we shall compare these results with the experimental data obtained
by the ARPES measurements.

The first step is to determine the position of the Fermi level $\varepsilon_{F}$ for the doping $\delta$
in the calculated band structure, then to find the corresponding Fermi surface
and finally to compare it with the empirical data. Since the discussion was
carried out in the hole picture, $1+ \delta$ holes are placed in the lowest band begining from
it's bottom at the ($\pi,\pi$) point, (i.e. the {\bf M} point of the
square unit cell), to the Fermi level $\varepsilon_{F}$, where $\varepsilon_{I}$ is
also possibly present for $t'>0$. As already mentioned in Sec.\ref{I}, the band structure
for $t'<0$ can be obtained from the one calculated for $t'>0$ by the reflection on
the {\bf k} plane. The result is that $\varepsilon_{F}$ falls then in the
$\varepsilon_{U}$ band alone, i.e. the band touching cannot occur close to the
$\varepsilon_{F}$. In any case, all three bands can be brought to the Fermi
level of $1+\delta$ holes by varying appropriately the tight-binding parameters.

Using in particular the regime separation summarized in Fig.\ref{6}, all the regimes of
the parameters are examined in this way, comparing the topology of the obtained Fermi
surfaces with the corresponding experimental data. The bending of the Fermi surfaces towards the
${\bf \Gamma}$ point requires that
$\Delta$ and $t'$ have the opposite signs for all the
considered materials. This is already apparent in the small $t'$ limit of Sec.\ref{II} and
generalizes to arbitrary $t'$. The values of the parameters vary among different materials
as will be discussed bellow. It is also appropriate to point out
that all the experimentally obtained Fermi surfaces are fitted here with only
two different parameters, e.g. with $t/2t'$ and $\Delta/2t'$. Two more parameters, e.g. t and
$\widetilde\varepsilon =(\varepsilon_{d}+2\varepsilon_{p})/3$ are needed to fit the
conducting                            
band(s) away
from $\varepsilon_{F}$.

\subsection{La$_{2-x}$Sr$_x$CuO$_4$}

\begin{figure}
\begin{center}
\epsfxsize=1.2in
\epsfbox{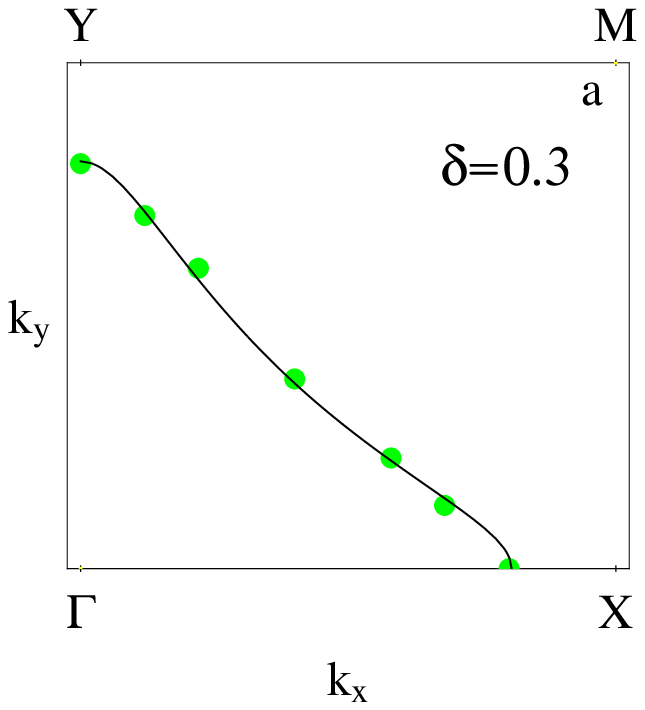}
\end{center}
\begin{figure}
\begin{center}
\epsfxsize=1.2in
\epsfbox{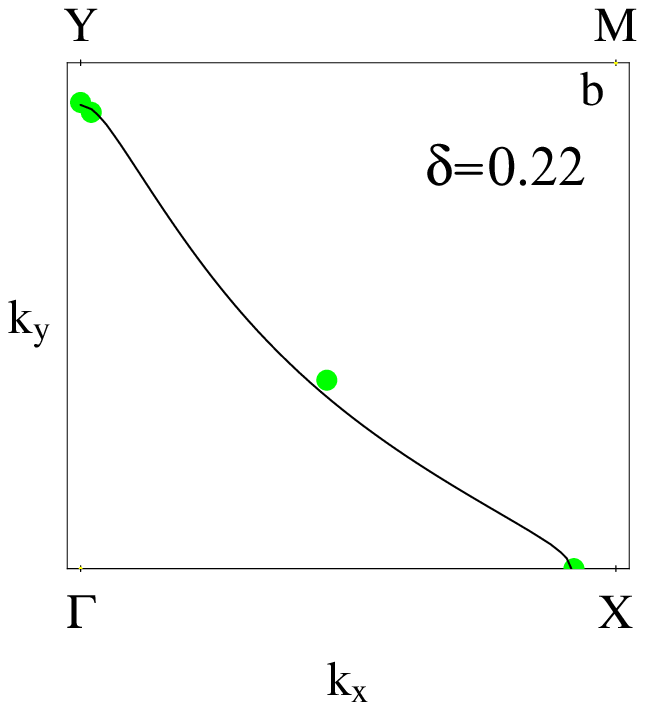}
\end{center}
\begin{figure}
\begin{center}
\epsfxsize=1.2in
\epsfbox{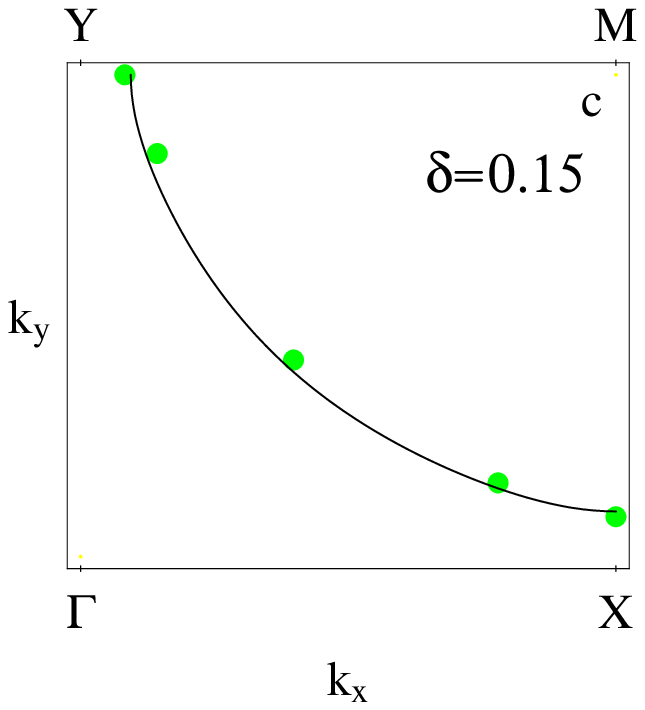}
\end{center}
\begin{figure}
\begin{center}
\epsfxsize=1.2in
\epsfbox{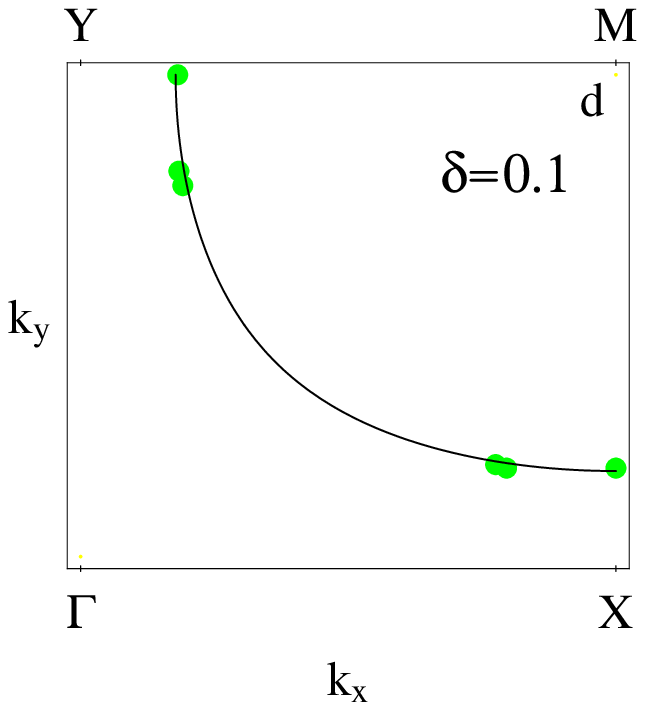}
\end{center}
\begin{figure}
\begin{center}
\epsfxsize=1.2in
\epsfbox{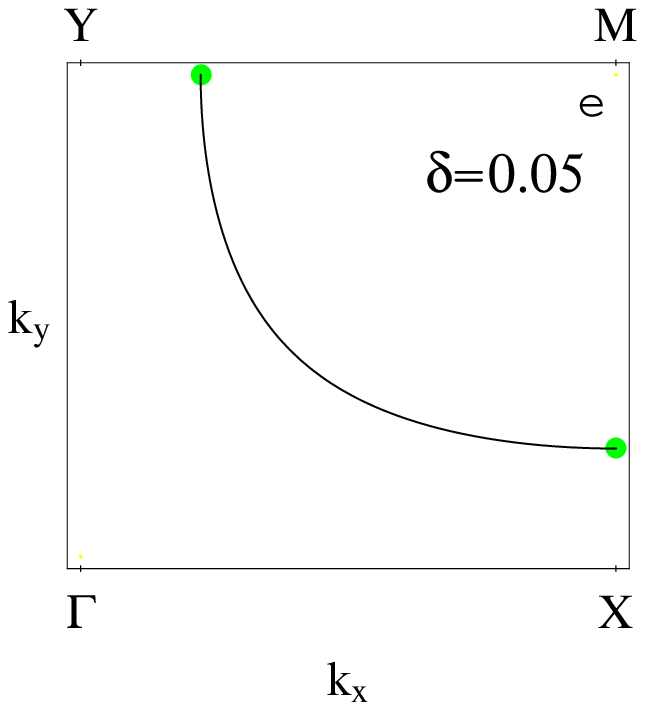}
\end{center}
\caption{
Experimentally measured Fermi surfaces for LSCO\protect\cite{Ino1,Ino2} (green dots) and the
three band model fits (solid). 
Fitting parameters are: 
\(
\Delta /2t'=-1.1,
t/2t'=0.56 (a),\\ 
\Delta /2t'=-0.75, 
t/2t'=0.5 (b),\\
\Delta /2t'=-0.5,
t/2t'=0.42 (c),\\ 
\Delta /2t'=-0.25, 
t/2t'=0.25 (d),\\ 
\Delta /2t'=-0.167, 
t/2t'=0.167 (e) 
\)
}
\label{fig10}
\end{figure}
\end{figure}
\end{figure}
\end{figure}       
\end{figure}

\begin{figure}
\begin{center}
\epsfxsize=1.2in
\epsfbox{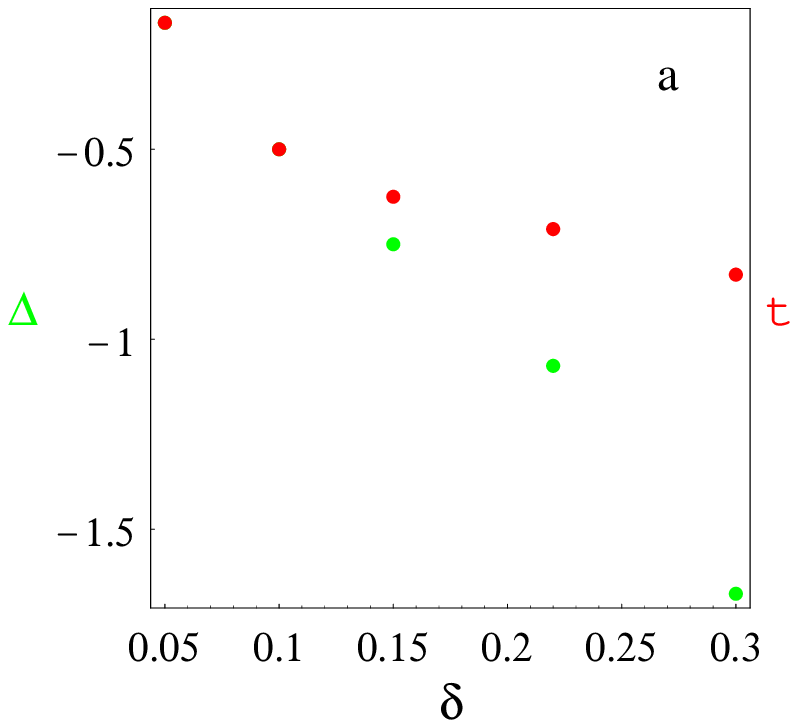}
\end{center}
\begin{figure}
\begin{center}
\epsfxsize=1.2in
\epsfbox{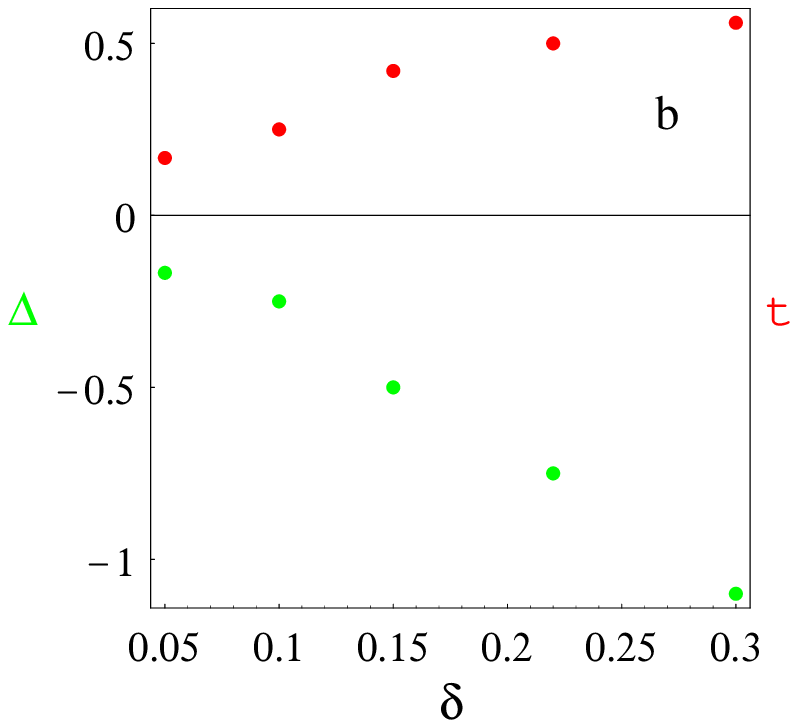}
\end{center}
\caption{Two sets of the fitting parameters with $\Delta>0,$ $t'<0$ (a) and $\Delta <0,$ $t'>0$
(b) in units 2$t'$, that
fit the experimenatally obtained Fermi surfaces\protect\cite {Ino1,Ino2} of LSCO, shown on Figs. 
\ref{fig10}a-e, 
equally well. For $\delta =$ 0.05 and $\delta =$0.1 on the Fig.\ref{fig11}a, $\Delta /2t'$ and
$t/2t'$ have
the same values.}
\label{fig11}
\end{figure}
\end{figure}

The recent ARPES measurements in LSCO \cite{Ino1,Ino2} give the evolution of the Fermi
surface in
the large range of dopings $\delta$. The two parameters fit of those data are
shown in Fig.\ref{fig10}a$-$e. Two sets of fitting parameters, given in Fig.\ref{fig11}, one
characterized by $t'>0$, $\Delta<0$ and the other by $t'<0$, $\Delta>0$, fit the
data equally well. It follows that the rigid band model is completely inadequate to describe the
evolution of the Fermi surface with $\delta$, because in both sets $\Delta/2t'$
and $t/2t'$ undergo large variations with $\delta$.

\vspace*{0.2cm}
\begin{multicols}{2}
\end{multicols}
\vspace*{0.2cm}

It is further important to note in Fig.\ref{fig11} that for both sets of parameters $\Delta /2t'$
and $t/2t'$ are
small for small $\delta$. This is best understood by noting that for $\Delta=0,t=0$ the Fermi
surface of the half-filled ($\delta=0$) band coincides with k$_x$, k$_y$ axes of the BZ, which in
the qualitative terms corresponds to the Figs.\ref{fig10}a$-$e. As already explained in Sections
\ref{II}
and \ref{III}, the energetic degeneracy of the ${\bf \Gamma X}$ line in the pure ($t=0$) oxygen
model
is related to the logarithmic rather than to the extended van Hove singularity in the density of
states.

\vspace*{0.2cm}
\begin{multicols}{2}
\end{multicols}
\vspace*{0.2cm}

\begin{figure}
\begin{center}
\epsfxsize=1.2in
\epsfbox{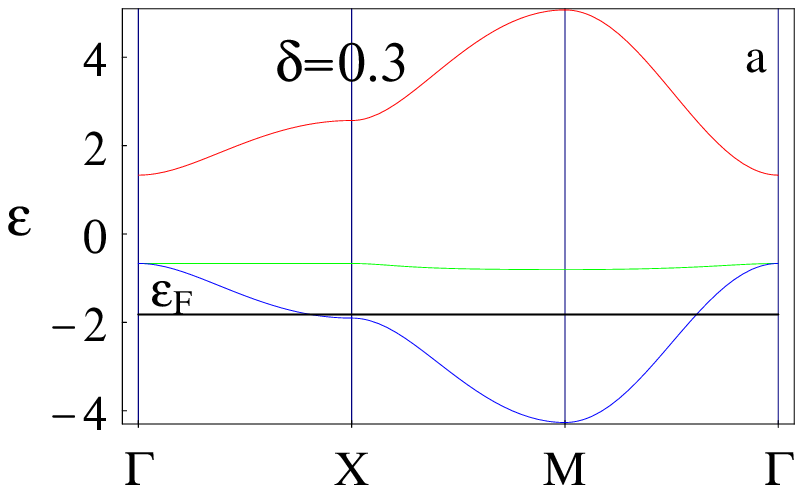}
\end{center}
\begin{figure}
\begin{center}
\epsfxsize=1.2in
\epsfbox{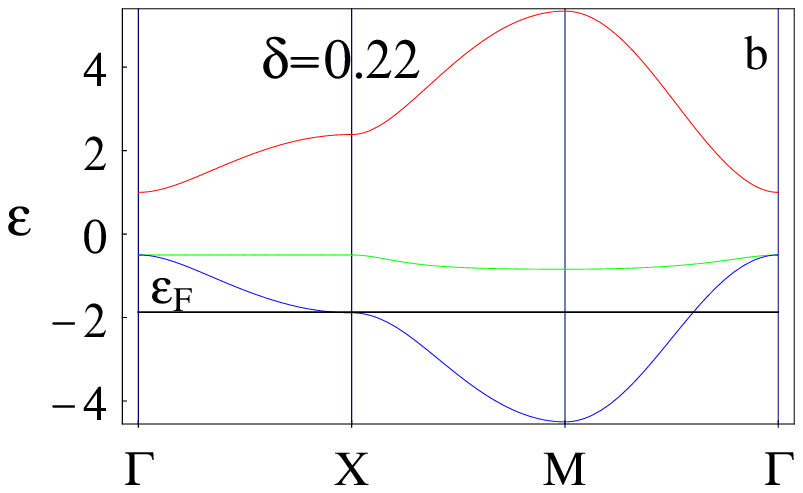}
\end{center}
\begin{figure}
\begin{center}
\epsfxsize=1.2in
\epsfbox{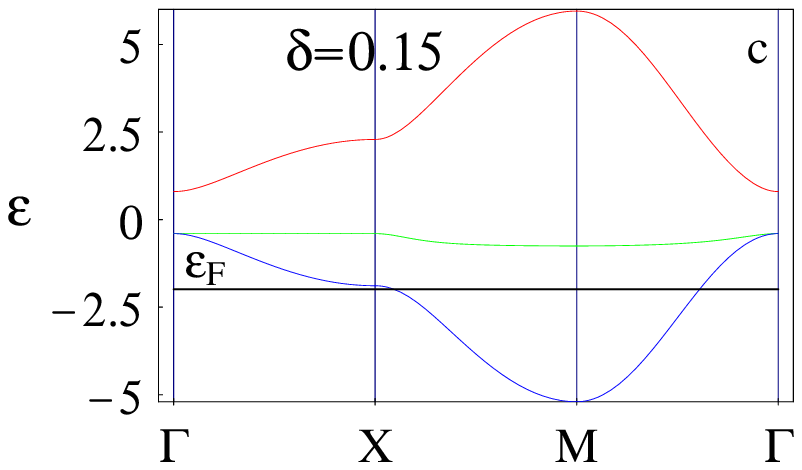}
\end{center}
\begin{figure}
\begin{center}
\epsfxsize=1.2in
\epsfbox{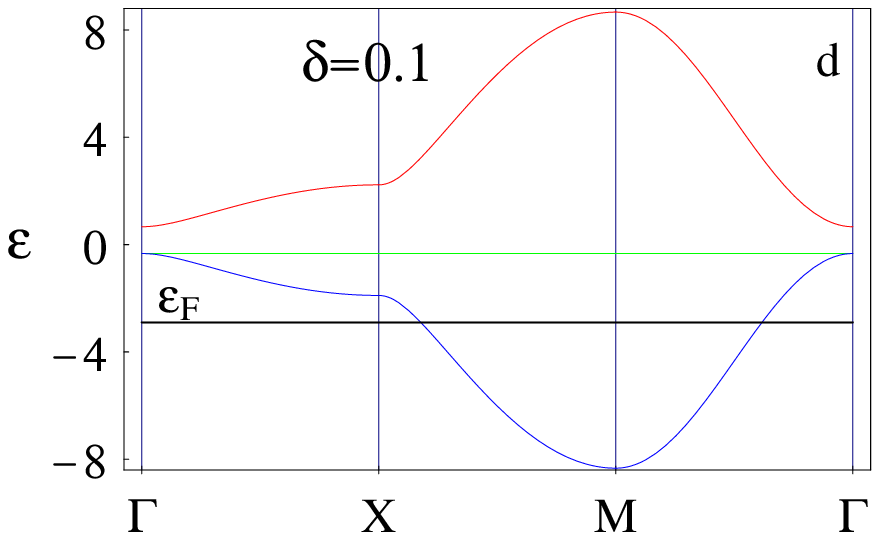}
\end{center}
\begin{figure}
\begin{center}
\epsfxsize=1.2in
\epsfbox{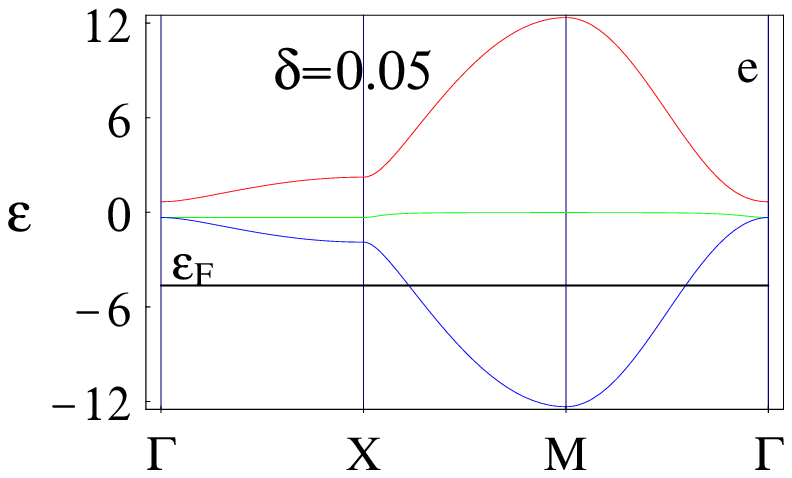}
\end{center}
\caption{Calculated three band structure obtained for the parameters of
Fig.\ref{fig10}, energy in units $t$.}
\label{fig12}
\end{figure}
\end{figure}
\end{figure}
\end{figure}
\end{figure}

The full band structure with $\Delta$ and $t$ small, but finite is shown in the
Fig.\ref{fig12} for $\Delta<0,t'>0$. In particular Figs.\ref{fig12}e and d illustrate that weak
dispersion is introduced by $t$ in the conducting band $\varepsilon_{L}$ along the ${\bf \Gamma X}$
line, i.e. the van Hove singularity at {\bf X}, although enhanced, is logarithmic. Fig.\ref{fig12}
also shows that for the values of parameters found in the case $\Delta<0, t'<0$ the band
$\varepsilon_{I}$ is rather far from the Fermi
level. For $\Delta>0, t'>0$, this holds independently on the values of the parameters.

\vspace*{0.2cm}
\begin{multicols}{2}
\end{multicols}
\vspace*{0.2cm}

The usual $t'=0$
mean field slave boson theory with large U, other interactions absent, allows small and even
negative
values of $\Delta$ \cite{Tutis}, but large variations of $\Delta$ and $t$ with the doping
$\delta$ are not obtained this way. The slave boson calculations with finite
$t'$ are under way for this reason and also because the previous results with finite $t'$ do not
cover the range of parameters which appeared to be relevant here.

\subsection{Bi$_2$Sr$_2$CaCu$_2$O$_8$}

As interesting feature of the ARPES results in Bi2212 taken in the normal state is the appearance of two
Fermi surfaces in the BZ \cite{Aebi,Ding,Mesot1,Fretwell,Borisenko}. There are points in
the 2d {\bf k} space in which two Fermi lines cross or touch. In the three band model
such points can occur only on the ${\bf \Gamma M}$ line, as shown in
Figs.\ref{fig7},\ref{fig8} and \ref{fig9},
whereas the experimantal crossing or touching points fall quite far from this line. It is
therefore more appropriate to interpret two observed Fermi lines as
belonging to the main band of the same topology as the one found in LSCO, and to
the shadow band obtained by the $(\pi,\pi)$ shift of the main band. It is likely
that this shift is produced by AF or by structural $(\pi,\pi)$
superlattice. The values of parameters which fit the main band are $\Delta/2t'=-0.0625$ and
$t/2t'=-0.1$, i.e. Bi2212 has even smaller values of $\Delta$ and $t$ than LSCO with
$\delta \approx 0.05$. However, the number of holes $1+\delta $ below Fermi level corresponds to
$\delta \approx 0.25$. Similar conclusion holds also for $\Delta<0,$ $t'>0$.

\begin{figure}
\begin{center}
\epsfxsize=1in
\epsfbox{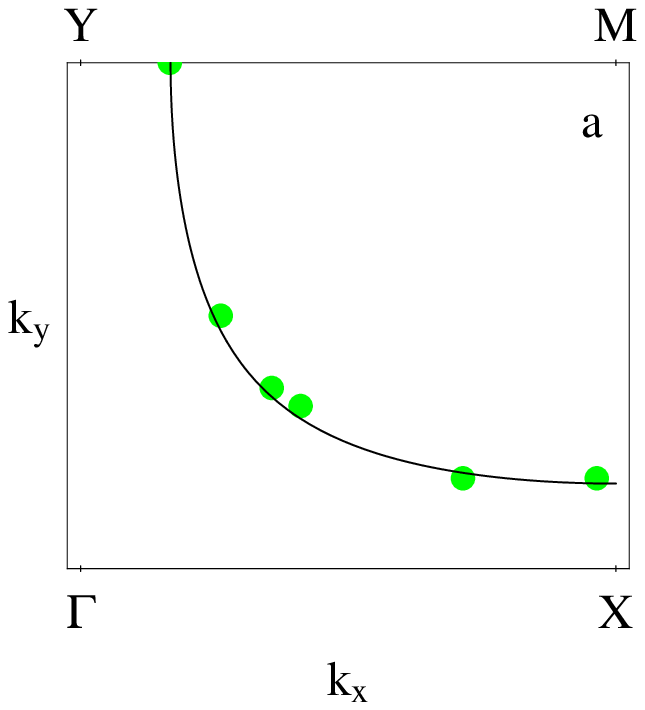}
\end{center}
\begin{figure}
\begin{center}
\epsfxsize=1in
\epsfbox{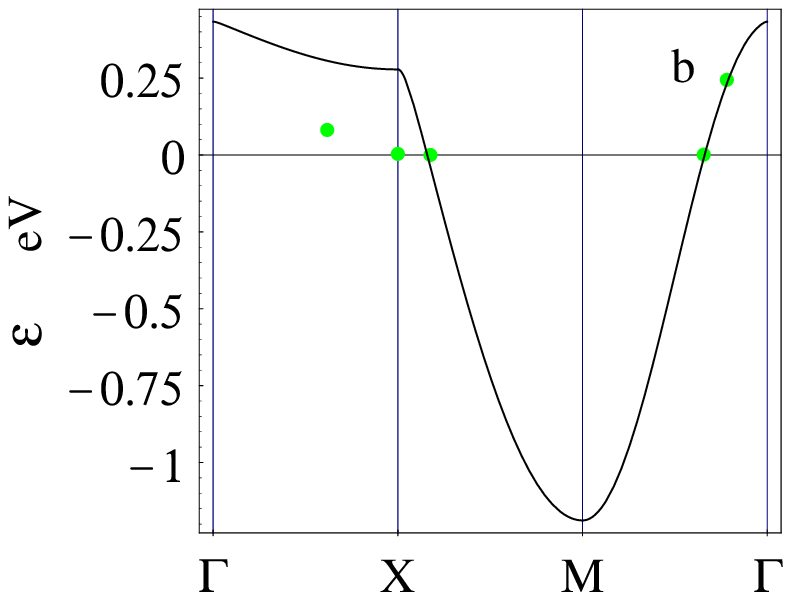}
\end{center}
\begin{figure}
\begin{center}
\epsfxsize=1.2in
\epsfbox{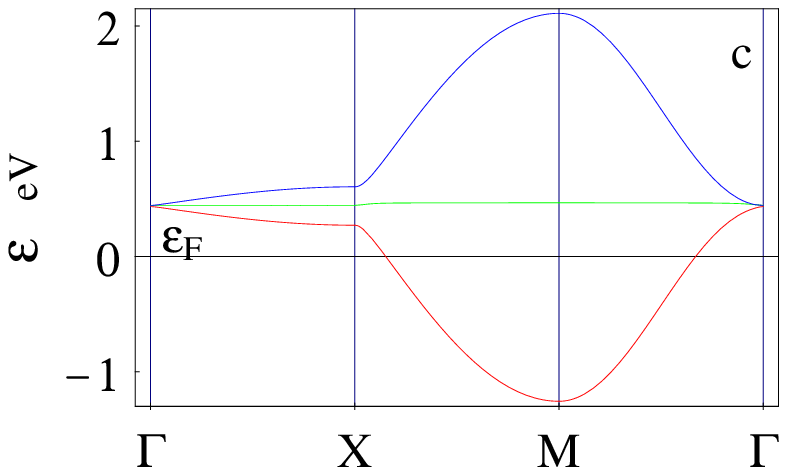}
\end{center}   
\caption{Experiments in Bi2212\protect\cite{Norman1} for the Fermi surface (a) and band
structure (b)(green dots), the three band model fits (solid) and the full three band structure (c) obtained
for the parameters $\Delta/2t'=$-0.0625, $t/2t'=$-0.1, $\widetilde\varepsilon=$0.44eV and
$t=$0.08eV }
\label{fig13}
\end{figure}
\end{figure}
\end{figure}

In addition to the Fermi surfaces the ARPES measurements in Bi2212 provide the
information about the states filled with electrons, i.e. about the empty hole
states \cite{Norman1}.  When the fit is extended to the energies away from the          
$\varepsilon_{F}$, $\widetilde\varepsilon$ and $t$ are determined. The best
fits to those measurements are shown in
Fig.\ref{fig13}b. The
idea
underlying these fits is that the Fermi liquid theory applies to the effects of
magnetic, structural or superconducting fluctuations, beyond the mean field
slave boson theory. The latter gives only the basic band structure, which was
discussed here. The Fermi liquid renormalisation of this structure by
fluctuations presumably vanish at
$\varepsilon_{F}$, i.e. the band theory is accurate at $\varepsilon_{F}$
\cite{SB1}, as it
is implicit from the fits of the Fermi surfaces, given in Figs.\ref{fig10} and
\ref{fig13}. The
departures of the experimental data from such fits are thus to be associated
with the fluctuation effects away from $\varepsilon_{F}$.

\vspace*{0.2cm}
\begin{multicols}{2}   
\end{multicols}
\vspace*{0.2cm}

The most prominent fluctuation effect defined this way is the flattening of the
measured dispersion along the ${\bf XM}$ line, compared to the band model
prediction (Fig.\ref{13}b). This flattening is understood here as the precursor of the gap
opening around the {\bf X} point, which is observed e.g. on doping Bi2212 with
Dy \cite{Marshall}. The precursor is not expected to be a simple pseudogap, which occurs only
when the adiabatic approximation can be applied to the relevant fluctuations.
Due to the small velocity of electrons, the adiabatic approximation fails in the
van Hove singularity, and the band states are conserved at the Fermi level,
although a part of them can be transferred to the pseudogap wings. This
theoretical prediction, which justifies the description of the Fermi surface by
the band theory, is borne out by the recent high resolution ARPES measurements
along the ${\bf\Gamma X}$ line \cite{Pavuna}. Those data show empirically that the states
close to $\varepsilon_{F}$ in the vicinity of the {\bf X} point are shared
between the peak at $\varepsilon_{F}$ and the pseudogap wings. The calculations
of the detailed {\bf k} dependances and the comparison with the experiments
is currently under way.

\subsection{YBa$_2$Cu$_3$O$_6.95$}

There are two planes per unit cell of Y123 which result in dimerisation of the
CuO$_2$ bands along the axis perpendicular to the planes. In addition, the CuO
chains of Y123 are conducting, which complicates further the analysis of the
observed band structure. It is however resonable to assume that the lowest of
the CuO$_2$ bands undergoes only the dispersionless dimerisation shift towards
lower energies.

\begin{figure}
\begin{center}
\epsfxsize=2in
\epsfbox{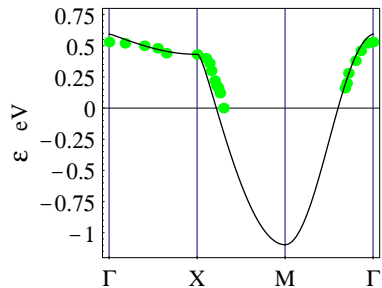}
\end{center}
\caption{Experimentally measured structure of the lowest band of Y123
\protect\cite{Shabel1} (green dots) and
the three band
model fits (solid) with parameters  $\Delta/2t'=$-0.0625, $t/2t'=$-0.1, $\widetilde\varepsilon=$0.58eV and $t=$0.08eV}
\label{fig14}
\end{figure}

The fit of the lowest band measured in Y123 \cite{Shabel1,Shabel2} with the three band model, shown
in Fig.\ref{fig14},
proceeds similarily to the cases of LSCO and Bi2212. The Fermi surface fit leads to
$\Delta/2t'$=-0.0625, $t/2t'$=-0.1 i.e.
to the values characteristic for Bi2212. It is interesting to note that the
band flattening observed in Bi2212 along the ${\bf XM}$ line does not occur in
Y123. If not a problem of the experimental resolution, this suggests that the nature of
fluctuations which affect the vicinity of
the {\bf X} point in the electron spectrum differs considerably between Bi2212
and Y123 and opens an interesting question which requires further investigation.

\vspace*{0.2cm}
\vspace*{0.2cm}

\section{Conclusion}
In contrast to the most of the many body theories of the high-T$_c$ superconductors, which are
attempting to construct the electron band structure starting from a given set of bare band and
interaction parameters, this paper assumes that the resulting average band structure can be
described with a few effective near-neighbor tight-binding parameters. After the detailed analysis,
which encompasses the rich structure of the three band model, it appears that the single band at
the Fermi level describes the Fermi surfaces in LSCO, Bi2212 and Y123 remarkably well. Only two
parameters are used in those fits, compared to the usual, at least four parameter fits of the
Fermi surfaces. The improvement is obtained by the use of the effective
tight-binding band structure which is non-linear in low order harmonics in contrast to the
Fourier fits. This indicates that the average band structure arises from the (large) short range
correlations. The departure of the predicted band behavior, measured in the vicinity the Fermi
level, is
used to determine the effects of the fluctuations on the small energy scales not included in the
average band structure. This way the
desired results of the many body theories are defined more clearly, they should end up with
the effective tight-binding parameters in the energy range between 1eV and 0.1eV, and with the
additional fluctuation effects on the energy scales of the order of 0.1eV to 0.01eV. The
corresponding space scales of the correlations involved into the average band structure and into
departures from it are related in analogous way. Large energy scales alone describe
the Fermi surface, what implies that the Fermi liquid theory is the good starting point in the
calculation of the fluctuation effects.

\section{Acknowledgment}
We would like to thank Prof. Jacques Friedel, Dr. Davor Pavuna, Dr. Eduard Tuti{\v s}, Dr. Denis
Sunko and Dr. Ivan Kup{\v c}i{\' c} for the stimulating discussions. This
work was supported by Croatian Ministry of Science under the project 119-204.

\end{multicols}
\end{document}